\documentclass[12pt]{article}
\usepackage{amsmath,amssymb,bm,graphicx}
\usepackage{cite}
\usepackage{color} 
\usepackage{hyperref}

\newcommand{\delslash}{\not \! \partial}

\begin{document}

\vskip 0.5 truecm

\begin{center}
{\Large{\bf Parity-doublet representation of Majorana fermions
and neutron oscillation}}
\end{center}
\vskip .5 truecm
\begin{center}
{\bf { Kazuo Fujikawa$^{1,2}$ and Anca Tureanu$^1$}}
\end{center}

\begin{center}
\vspace*{0.4cm} 
{\it {$^1$Department of Physics, University of Helsinki, P.O.Box 64, 
\\FIN-00014 Helsinki,
Finland\\
$^2$Quantum Hadron Physics Laboratory, RIKEN Nishina Center,\\
Wako 351-0198, Japan
}}
\end{center}


\begin{abstract}
We present a parity-doublet theorem for the representation of the intrinsic parity of Majorana fermions, which is expected to be useful  also in condensed matter physics, and it is illustrated to provide a criterion of neutron-antineutron oscillation in a BCS-like effective theory with $\Delta B=2$  baryon number violating terms.  The CP violation in the present effective theory causes  no direct CP violating effects in the oscillation itself, which is demonstrated by the exact solution, although it influences the neutron electric dipole moment in the leading order of small $\Delta B=2$ parameters.   An  analogue of Bogoliubov transformation, which preserves P and CP, is crucial in the analysis. 
\end{abstract}
 
\section{Introduction}

The Majorana fermions received much attention recently not only in particle physics~\cite{bilenky} but also in condensed matter 
physics~\cite{condensed-matter}. The Majorana fermions are defined by the condition 
$\psi(x)=C\bar{\psi}^{T}(x)=\psi^{c}(x)$
where $C=i\gamma^{2}\gamma^{0}$ stands for the charge conjugation 
matrix~\cite{bjorken}. 

We start with a neutral Dirac fermion $n(x)$ and define the combinations
$\psi_{\pm}(x)=\frac{1}{\sqrt{2}}[n(x)\pm n^{c}(x)]$,
which satisfy
$\psi^{c}_{\pm}(x)=\pm\psi_{\pm}(x)$,
showing that $\psi_{+}(x)$ and $\psi_{-}(x)$ are Majorana fields. We treat the 
fermion with $\psi^{c}_{-}(x)=-\psi_{-}(x)$ also as a Majorana fermion.
It is well-known~\cite{bjorken,weinberg} that,  in theories where the fermion number is conserved, the discrete symmetry such as parity can generally be defined  with an arbitrary phase freedom $\delta$,
$n(x)\rightarrow e^{i\delta}\gamma^{0}n(t,-\vec{x})$.
As working definitions of C, P and T even in a  theory with fermion number violation, we use the conventional definitions given in~\cite{bjorken}. 
One can then confirm that the parity operation with vanishing phase $\delta=0$, which is the most common definition~\cite{bjorken,weinberg} and called  "$\gamma^{0}$-parity" in the following when we specify it more precisely, 
namely,
\begin{eqnarray}
n(x)\rightarrow \gamma^{0}n(t,-\vec{x}),\ \ \ n^{c}(x)\rightarrow -\gamma^{0}n^{c}(t,-\vec{x})
\end{eqnarray}
that satisfy ${\rm P}^{2}=1$,
leads to a {\em doublet representation} $\{\psi_{+}(x), \psi_{-}(x) \}$,
\begin{eqnarray}\label{1}
\psi_{\pm}(x)\rightarrow \gamma^{0}\psi_{\mp}(t,-\vec{x}).
\end{eqnarray}
Only when the two fermions $\psi_{\pm}(x)$ are degenerate, this doublet representation is consistent with dynamics~\cite{weinberg}.  The mass splitting in $\psi_{\pm}(x)$ inevitably breaks the $\gamma^{0}$-parity as a symmetry of the Lagrangian.

On the other hand, one confirms that parity for an {\em isolated single} Majorana fermion is consistently defined only by "$i\gamma^{0}$-parity" with $\delta=\pi/2$,
\begin{eqnarray}
n(x)\rightarrow i\gamma^{0}n(t,-\vec{x}), \ \ \ n^{c}(x)\rightarrow i\gamma^{0}n^{c}(t,-\vec{x}),
\end{eqnarray}
namely by~\cite{weinberg},
\begin{eqnarray}\label{2}
\psi_{\pm}(x)\rightarrow i\gamma^{0}\psi_{\pm}(t,-\vec{x}),
\end{eqnarray}
which is consistent with the reality of $\psi_{\pm}(x)$ in the  Majorana representation where $\gamma^{0}$ is hermitian but purely imaginary. 
The phase freedom $\delta$ is thus fixed by the Majorana condition and ${\rm P}^{2}=-1$.  This transformation rule \eqref{2} by itself does not tell the presence or absence of the mass splitting of $\psi_{\pm}(x)$.

The intrinsic parity of Majorana fermions, in particular, the parity-doublet
theorem stated below, are shown to play main roles in the discussion of neutron-antineutron 
oscillation~\cite{mohapatra, mohapatra2}, which received attention 
recently~\cite{berezhiani,ft, nelson,gardner}.

\section{Effective $\Delta B=2$ Lagrangian for the neutron}

In the analysis of possible baryon number violation and neutron oscillation, one can study essential aspects by analyzing the quadratic effective hermitian Lagrangian with general $\Delta B=2$ terms  added,
\begin{eqnarray}\label{3}
{\cal L}&=&\overline{n}(x)i\gamma^{\mu}\partial_{\mu}n(x) - m\overline{n}(x)n(x)\nonumber\\
&-&\frac{i}{2}\epsilon_{1}[e^{i\alpha}n^{T}(x)Cn(x) - e^{-i\alpha}\overline{n}(x)C\overline{n}^{T}(x)]\nonumber\\
&-&\frac{i}{2}\epsilon_{5}[n^{T}(x)C\gamma_{5}n(x) + \overline{n}(x)C\gamma_{5}\overline{n}^{T}(x)],
\end{eqnarray}
where $m$, $\epsilon_{1}$, $\epsilon_{5}$ and $\alpha$ are real parameters. The most general quadratic hermitian Lagrangian is written in the form \eqref{3} using the phase freedom of $n(x)\rightarrow n(x)=e^{i\beta}n^{\prime}(x)$; under this change of naming the field, the physical quantities
in \eqref{3} such as mass eigenvalues are obviously invariant. But once one defines $n^{c}(x)\equiv C\bar{n}^{T}$, we have $n^{c}(x)=e^{-i\beta}{n^{\prime}}^{c}(x)$ and thus the C and CP properties of the solutions of the Lagrangian \eqref{3} are changed. 
The first $\Delta B=2$ term with real $\epsilon_{1}$ breaks the $\gamma^{0}$-parity while the second $\Delta B=2$ term with real $\epsilon_{5}$ preserves $\gamma^{0}$-parity.  The  term with $\epsilon_{5}$ written in the variables in \eqref{3} preserves C and CP. (In contrast, the first $\Delta B=2$ term preserves $i\gamma^{0}$-parity while the second $\Delta B=2$ term breaks $i\gamma^{0}$-parity.) An analogy of neutron oscillation in \eqref{3} to BCS theory has been emphasized at the early stage of the study of neutron oscillation~\cite{chang}. 

We assume that the baryon number of the neutron in our effective theory is fixed by 
strong interactions and thus a parity violating $\Delta B=0$ mass term $-\delta m^{\prime}\bar{n}i\gamma_{5}n$ induced by the QCD $\theta$-vacuum, for example, should be added to a possible 
parity violating mass term induced by $\Delta B=2$ interactions discussed later.  

The model \eqref{3} with $\alpha=0$, which preserves CP, is related to the single-flavor neutrino model in Ref. \cite{bilenky} if one replaces $n(x)$ by $\nu(x)$ and suitably adjusts the phase of the neutrino in the latter model. The mass eigenvalues are given by
\begin{eqnarray}\label{4}
M_{\pm}=\sqrt{m^{2}+\epsilon_{5}^{2}}\pm \epsilon_{1}.
\end{eqnarray}

For the purpose of the analysis of CP violation later, the hermitian Lagrangian \eqref{3} with $\alpha=-\pi/2$ is interesting. For this choice,
the second  term with $\epsilon_{1}$ breaks $\gamma^{0}$-parity and CP, although C is a good symmetry of the total Lagrangian.
The Lagrangian with $\alpha=-\pi/2$ is exactly solved in terms of the solutions of 
\begin{eqnarray}
(i\delslash-M_{\pm})\psi_{\pm}(x)=0,
\end{eqnarray}
with mass eigenvalues 
\begin{eqnarray}\label{5}
M_{\pm}=\sqrt{(m\pm\epsilon_{1})^{2}+\epsilon^{2}_{5}}.
\end{eqnarray}
The neutron fields are then given by
\begin{eqnarray}\label{6}
\frac{1}{\sqrt{2}}\left(\begin{array}{c}
            n(x)+n^{c}(x)\\
            n(x)-n^{c}(x)
            \end{array}\right)
&=& e^{-i\Theta\gamma_{5}}e^{-i\tau_{3}\gamma_{5}\bar{\Theta}}
            \left(\begin{array}{c}
            \psi_{+}(x)\\
            \psi_{-}(x)
            \end{array}\right),
\end{eqnarray}
where the chiral phase factors are defined by
\begin{eqnarray}
M_{\pm}e^{2i\tilde{\theta}_{\pm}\gamma_{5}}=m\pm\epsilon_{1}\pm i\epsilon_{5}\gamma_{5},
\end{eqnarray}
and $\Theta\equiv \frac{1}{2}(\tilde{\theta}_{+}+\tilde{\theta}_{-})$ and $\bar{\Theta}\equiv \frac{1}{2}(\tilde{\theta}_{+}-\tilde{\theta}_{-})$.
The charge conjugation properties of $\psi_{\pm}$ are~\cite{note}
\begin{eqnarray}\label{7}
\psi^{c}_{+}(x)=\psi_{+}(x), \ \ \ \psi^{c}_{-}(x)=-\psi_{-}(x),
\end{eqnarray}
and thus Majorana fermions, which are consistent with the C transformation of $n$ and $n^{c}$ in \eqref{6} and also consistent with the C-invariance of  \eqref{3} (with $\alpha=-\pi/2$).
One can confirm that the doublet representation of parity
\begin{eqnarray}\label{8}
\psi_{\pm}(x)\rightarrow \gamma^{0}\psi_{\mp}(t,-\vec{x})
\end{eqnarray}
does not induce the $\gamma^{0}$-parity transformation of $n$ and $n^{c}$ for $\Theta\neq 0$ in \eqref{6}, reflecting the P (and thus CP) violation for $\epsilon_{1}\epsilon_{5}\neq 0$ (for $\epsilon_{1}\epsilon_{5}= 0$, $\Theta=0$) in addition to the dynamical P breaking $M_{+}\neq M_{-}$ for $\epsilon_{1}\neq 0$ in \eqref{3}.  

It is confirmed that the CP violation in the Lagrangian \eqref{3} (with $\alpha\neq 0$) is not eliminated by any phase choice of the neutron field $n(x)\rightarrow e^{i\beta}n(x)$ for real $\epsilon_{1}\neq 0$ and $\epsilon_{5}\neq 0$, and in this sense it is {\em intrinsic}; $\epsilon_{1}\epsilon_{5}\neq 0$ is a necessary condition of CP violation, and parity is inevitably broken.

\section{Parity-doublet theorem}
The presence of the neutron oscillation $P(n\rightarrow \bar{n})\propto \sin^{2}((\Delta M/2) t)$ implies the mass splitting of auxiliary Majorana-type fermions as in \eqref{4} and \eqref{5}. Phenomenologically, one thus observes that the $\gamma^{0}$-parity violation by $\epsilon_{1}\neq 0$ in \eqref{3}, which gives rise to 
$M_{+}\neq M_{-}$, is the necessary condition 
of neutron oscillation. 
The observation of the neutron oscillation thus implies the dynamical inconsistency of the doublet representation of $\gamma^{0}$-parity. In contrast, the failure (or success) of $i\gamma^{0}$-parity does not tell the presence or absence of neutron oscillation.

In the following we shall make more precise the role of $\gamma^{0}$-parity and show that the main features of the solutions of the Lagrangian  \eqref{3} can be obtained from general symmetry considerations. We first rewrite \eqref{3} in the notation of $n(x)$ and $n^{c}(x)$ as 
\begin{eqnarray}\label{3'}
S(n, n^{c},\epsilon_{1})
&=&
\int d^{4}x\Big\{\frac{1}{2}\overline{n}(x)[i\gamma^{\mu}\partial_{\mu} - m]n(x)+\frac{1}{2}\overline{n^{c}}(x)[i\gamma^{\mu}\partial_{\mu}- m]n^{c}(x)\nonumber\\
&&-\frac{i}{2}\epsilon_{1}[e^{i\alpha}\overline{n^{c}}(x)n(x) - e^{-i\alpha}\overline{n}(x)n^{c}(x)]\nonumber\\
&&-\frac{i}{2}\epsilon_{5}[\overline{n^{c}}(x)\gamma_{5}n(x) + \overline{n}(x)\gamma_{5}n^{c}(x)]\Big\}.
\end{eqnarray}
The CP (or T) symmetry of the effective Lagrangian is fixed by the choice of $\alpha$ in \eqref{3'}, therefore we start with an arbitrary $\alpha$. The quadratic Lagrangian can be diagonalized and the general solution corresponding to the eigenvalue $M(\epsilon_{1})$, 
\begin{eqnarray}\label{gen_sol}
[i\gamma^{\mu}\partial_{\mu}-M(\epsilon_{1})]\psi_{+}(x; \epsilon_{1})=0,
\end{eqnarray}
is written as 
\begin{eqnarray}\label{10}
\psi_{+}(x; \epsilon_{1})\equiv c_{1}n(x)+c_{2}\gamma_{5}n(x)+c_{3}n^{c}(x)+c_{4}\gamma_{5}n^{c}(x),
\end{eqnarray}
with suitable complex constants $\{c_{j}(\epsilon_{1})\}$ that generally depend on $\epsilon_{1}$ (as well as on $m$, $\epsilon_{5}$ and $\alpha$, but these parameters  do not influence the transformation property under the $\gamma^{0}$-parity). One can easily ascertain that the action \eqref{3'} is invariant under the $\gamma^{0}$-parity transformation of $n(x)$ and $n^{c}(x)$ combined with the inversion $\epsilon_{1}\rightarrow -\epsilon_{1}$, namely,
\begin{eqnarray}\label{broken_parity_rel}
S(n, n^{c},\epsilon_{1})
=S(n^{p}, (n^{c})^{p},-\epsilon_{1}).
\end{eqnarray}
Consequently, if one performs a $\gamma_0$-parity transformation and the inversion $\epsilon_{1}\rightarrow -\epsilon_{1}$ on the solution \eqref{gen_sol}, one finds a solution of $S(n^{p}, (n^{c})^{p},-\epsilon_{1})$:
\begin{eqnarray}\label{11}
\psi_{+}^{p}(x; -\epsilon_{1})&=&[c_{1}(-\epsilon_{1})n^{p}(x)+c_{2}(-\epsilon_{1})\gamma_{5}n^{p}(x)+c_{3}(-\epsilon_{1})(n^{c}(x))^{p}+c_{4}(-\epsilon_{1})\gamma_{5}(n^{c}(x))^{p}]\nonumber\\
&=&\gamma^{0}[c_{1}(-\epsilon_{1})n(t,-\vec{x})-c_{2}(-\epsilon_{1})\gamma_{5}n(t,-\vec{x})\nonumber\\
&&-c_{3}(-\epsilon_{1})n^{c}(t,-\vec{x})+c_{4}(-\epsilon_{1})\gamma_{5}n^{c}(t,-\vec{x})]\nonumber\\
&\equiv& \gamma^{0}\psi_{-}(t,-\vec{x}; -\epsilon_{1}),
\end{eqnarray}
which satisfies (as it will be justified below using the Ward--Takahashi identity)
\begin{eqnarray}\label{11'}
[i\gamma^{\mu}\partial_{\mu}-M(-\epsilon_{1})]\psi_{+}^{p}(x; -\epsilon_{1})=0.
\end{eqnarray}
This relation implies
\begin{eqnarray}
&&\gamma^{0}[i\gamma^{\mu}\partial_{\mu}-M(-\epsilon_{1})]\psi_{+}^{p}(x; -\epsilon_{1})\nonumber\\
&&=\gamma^{0}[i\gamma^{\mu}\partial_{\mu}-M(-\epsilon_{1})]\gamma^{0}\psi_{-}(t,-\vec{x}; -\epsilon_{1})=0.
\end{eqnarray}
Thus we have found a solution of 
$S(n^{p}, (n^{c})^{p},-\epsilon_{1})=S(n, n^{c},\epsilon_{1})$, corresponding to the eigenvalue $M(-\epsilon_1)$:
\begin{eqnarray}\label{xx}
[i\gamma^{\mu}\partial_{\mu}-M(-\epsilon_{1})]\psi_{-}(x; -\epsilon_{1})=0.
\end{eqnarray}
The $\gamma^{0}$-parity transformation maps a solution
$\psi_{+}(x; \epsilon_{1})$ with mass $M(\epsilon_1)$ in \eqref{gen_sol} to another solution $\psi_{-}(x; -\epsilon_{1})$ with mass $M(-\epsilon_1)$ in \eqref{xx} as is suggested by the broken parity symmetry relation in \eqref{broken_parity_rel}.
 
We return now to the justification of the mass eigenvalue $M(- \epsilon_1)$ in eq. \eqref{11'}. We note that the solution 
$[i\gamma^{\mu}\partial_{\mu}-M(-\epsilon_{1})]\psi_{+}^{p}(x; -\epsilon_{1})=0$ of the action $S(n^{p},(n^{c})^{p},-\epsilon_{1})$ in \eqref{11'} and the solution $[i\gamma^{\mu}\partial_{\mu}-M(-\epsilon_{1})]\psi_{+}(x; -\epsilon_{1})=0$ of the action $S(n,n^{c},-\epsilon_{1})$ share the same mass eigenvalue since one can regard $n\rightarrow n^{p}$ and $n^{c}\rightarrow (n^{c})^{p}$ as a formal re-naming of field variables. Remark, however, that $S(n,n^{c},-\epsilon_{1})\neq S(n,n^{c},\epsilon_{1})$ while $S(n^{p},(n^{c})^{p},-\epsilon_{1})=S(n,n^{c},\epsilon_{1})$. 
A justification of \eqref{11'} is given by a Ward--Takahashi identity
for broken parity symmetry in the path integral by starting with 
\begin{eqnarray}\label{WT}
&&\langle T^{\star} n^{p}(x)\overline{(n^{c})^{p}}(y)\rangle
{\Big|_{(\epsilon_{1})}}=\int {\cal D}n{\cal D}n^{c} n^{p}(x)\overline{(n^{c})^{p}}(y)e^{iS(n,n^{c},\epsilon_{1})}\nonumber\\
&&=\int {\cal D}n^{p}{\cal D}(n^{c})^{p} n^{p}(x)\overline{(n^{c})^{p}}(y)e^{iS(n^{p},(n^{c})^{p},-\epsilon_{1})}
\end{eqnarray}
where $T^{\star}$ stands for time-ordering, and we used the parity invariance of the path integral measure ${\cal D}n^{p}{\cal D}(n^{c})^{p}={\cal D}n{\cal D}n^{c}$ and the broken parity relation \eqref{broken_parity_rel}, $S(n^{p},(n^{c})^{p},-\epsilon_{1})=S(n,n^{c},\epsilon_{1})$. The last path integral in \eqref{WT} is identical to the path integral in
\begin{eqnarray}
\langle T^{\star} n(x)\overline{n^{c}}(y)\rangle\Big|_{(-\epsilon_{1})}=\int {\cal D}n{\cal D}n^{c} n(x)\overline{n^{c}}(y)e^{iS(n,n^{c},-\epsilon_{1})},
\end{eqnarray}
using the re-naming of path integral variables. We thus conclude
that $\langle T^{\star} n^{p}(x)\overline{(n^{c})^{p}}(y)\rangle
\Big|_{(\epsilon_{1})}$ defined by $S(n,n^{c},\epsilon_{1})$ agrees with $\langle T^{\star} n(x)\overline{n^{c}}(y)\rangle\Big|_{(-\epsilon_{1})}$ defined by $S(n,n^{c},-\epsilon_{1})$. This relation holds for other combinations of fields such as $\langle T^{\star} n^{p}(x)\overline{n^{p}}(y)\rangle$ also, and thus we conclude that
\begin{eqnarray}
\langle T^{\star}\psi_{+}^{p}(x; -\epsilon_{1})\overline{\psi_{+}^{p}}(y; -\epsilon_{1})\rangle
\end{eqnarray}
for the action $S(n,n^{c},\epsilon_{1})=S(n^{p},(n^{c})^{p},-\epsilon_{1})$ agrees with 
\begin{eqnarray}
\langle T^{\star}\psi_{+}(x; -\epsilon_{1})\overline{\psi_{+}}(y; -\epsilon_{1})\rangle
\end{eqnarray}
for the action $S(n,n^{c},-\epsilon_{1})$, i.e. they have the same pole mass, $M(-\epsilon_1)$.
This provides a justification of the mass eigenvalue in \eqref{11'}.

A similar analysis in the inverse direction starting with the solution $\psi_{-}(x; -\epsilon_{1})$ in \eqref{xx} of $S(n^{p}, (n^{c})^{p},-\epsilon_{1})$ leads to a solution $\psi_{+}(t,-\vec{x}; \epsilon_{1})$ of $S(n, (n^{c}),\epsilon_{1})$ with mass $M(\epsilon_{1})$. We have thus established the $\gamma^{0}$-parity doublet representation $\{\psi_{+}(x; \epsilon_{1}), \psi_{-}(x; -\epsilon_{1})\}$
of the solutions of $S(n, n^{c},\epsilon_{1})$,
\begin{eqnarray}\label{yy}
&&\psi_{+}(x; \epsilon_{1})\xrightarrow{P} \gamma^{0}\psi_{-}(t,-\vec{x}; -\epsilon_{1}),\nonumber\\
&&\psi_{-}(x; -\epsilon_{1})\xrightarrow{P}\gamma^{0}\psi_{+}(t,-\vec{x}; \epsilon_{1}),
\end{eqnarray}
which satisfy ${\rm P}^{2}=1$ ({\em parity-doublet theorem}). This representation is valid irrespective of whether the $\gamma^0$-parity is conserved or not. The $\gamma^{0}$-parity violation by $\epsilon_{1}\neq 0$ in \eqref{3'} is a necessary condition of neutron oscillation which requires  $M_{+}=M(\epsilon_{1})\neq M(-\epsilon_{1})=M_{-}$, in which case the doublet representation is dynamically inconsistent.  

A more detailed specification of the solutions is possible if one assumes some symmetry of the Lagrangian. For example, good C in \eqref{3} and \eqref{3'} with $\alpha=-\pi/2$ implies the relation $\psi_{+}^{c}=C\bar{\psi_{+}}^{T}$, where the left-hand side $\psi_{+}^{c}$ is evaluated in terms of $n$ and $n^{c}$ by a unitary C-transformation,
\begin{eqnarray}
\psi^c_{+}(x; \epsilon_{1})= c_{1}n^c(x)+c_{2}\gamma_{5}n^c(x)+c_{3}n(x)+c_{4}\gamma_{5}n(x),
\end{eqnarray}
while the right-hand side is evaluated directly from $\psi_{+}$,
\begin{eqnarray}
C\bar{\psi_{+}}^{T}= i\gamma^2\left[c_{1}n(x)+c_{2}\gamma_{5}n(x)+c_{3}n^{c}(x)+c_{4}\gamma_{5}n^{c}(x)\right]^*,
\end{eqnarray}
and one obtains
\begin{eqnarray}
\psi_{+}(x)=c_{1}n+ic_{2}\gamma_{5}n+c_{3}n^{c}+ic_{4}\gamma_{5}n^{c}
\end{eqnarray}
as a general expansion analogous to \eqref{10} where all the coefficients are now real. The condition $\psi^{c}_{+}=\psi_{+}$ and the parity doublet condition $\psi_{+}\rightarrow \gamma^{0}\psi_{-}$ in \eqref{yy} then fix
the general forms 
\begin{eqnarray}
\psi_{+}&=&(c_{1}(\epsilon_{1})+ic_{2}(\epsilon_{1})\gamma_{5})(n+n^{c}),\nonumber\\
\psi_{-}&=&(c_{1}(-\epsilon_{1})-ic_{2}(-\epsilon_{1})\gamma_{5})(n-n^{c}),
\end{eqnarray}
and the last expression $\psi_{-}$ also satisfies the condition $\psi^{c}_{-}=-\psi_{-}$. This is precisely the structure obtained earlier by direct calculations in eq. \eqref{6}.
Thus, we can derive the general features of the exact solution in \eqref{6} without solving explicitly the equations of motion, but just by using the $\gamma^{0}$-parity.  

The absence of the $\gamma^{0}$-parity violating $\epsilon_{1}$ term implies the absence of the conventional neutron oscillation, because of the mass degeneracy of the two solutions in \eqref{yy}, despite the presence of $\epsilon_{5}$ term with $\Delta B=2$, which breaks generally defined parity  in \eqref{3'}. We discuss what happens in this case by setting $\epsilon_{1}=0$ in \eqref{3'} which preserves C and P.
 The solution of this Lagrangian with $\epsilon_{1}=0$ is given by \eqref{6},
\begin{eqnarray}\label{12}
\left(\begin{array}{c}
            n(x)\\
            n^{c}(x)
            \end{array}\right)
&=& \left(\begin{array}{c}
            \cos\phi N_{+}(x)-i\gamma_{5}\sin\phi N_{-}(x)\\
            \cos\phi N_{-}(x)-i\gamma_{5}\sin\phi N_{+}(x)
            \end{array}\right),
\end{eqnarray}
but now with
\begin{eqnarray}
N_{\pm}(x)=[\psi_{+}(x)\pm \psi_{-}(x)]/\sqrt{2},
\end{eqnarray}
and 
\begin{eqnarray}\label{13}
\sin2\phi\equiv \epsilon_{5}/\sqrt{m^{2}+\epsilon_{5}^{2}},
\end{eqnarray}
which satisfy
$N_{\pm}^{c}(x)=N_{\mp}(x)$ and $N^{p}_{\pm}(x)=\pm \gamma^{0}N_{\pm}(t,-\vec{x})$ using \eqref{7} and \eqref{8}. The fields $N_{\pm}(x)$ have the degenerate mass 
\begin{eqnarray}\label{14}
M=\sqrt{m^{2}+\epsilon_{5}^{2}}.
\end{eqnarray}
In the notation of \eqref{10} and \eqref{yy}, the solution with good  $\gamma^{0}$-parity gives a degenerate pair of Majorana fermions or equivalently a Dirac fermion $N_{+}(x)$, and good C implies the doublet representation $\{N_{+}(x), N_{+}^{c}(x)\}$.

When one generates the neutron, one obtains 
\begin{eqnarray}
n(x)=\cos\phi N_{+}(x)-i\gamma_{5}\sin\phi N^{c}_{+}(x)
\end{eqnarray}
in \eqref{12}, but no oscillation in the conventional sense due to the mass degeneracy and thus it may appear that  there is no physical effects. But $n(x)$ and $n^{c}(x)$ are not orthogonal in the sense 
\begin{eqnarray}\label{15}
\langle T^{\star} n^{c}(x)\bar{n}(y)\rangle
= \int\frac{d^{4}p}{(2\pi)^{4}}\frac{\gamma_{5}M\sin2\phi}{p^{2}-M^{2}+i\epsilon}e^{-ip(x-y)},
\end{eqnarray}
which shows that $n(x)$ decays through $n\rightarrow p+e+\bar{\nu}_{e}$ or $n^{c}\rightarrow \bar{p}+e^{+}+\nu_{e}$, and $n(x)$ annihilates when it collides with ordinary matter containing the neutron.
The implication of the absence of oscillation (with $\Delta B=2$ terms present) is the absence of "bunching effect" in the sense that one would observe predominantly $n^{c}(x)$
starting with $n(x)$ when observed at a proper moment in the presence of oscillation.  

In analogy of the neutron oscillation to BCS theory~\cite{chang}, we note that the relation \eqref{12} is precisely a 
Lorentz invariant version of the Bogoliubov transformation~\cite{FT} which diagonalizes the Lagrangian (with $\epsilon_{1}=0$) by preserving the anti-commutation relations 
\begin{eqnarray}
\{n(t,\vec{x}), n^{c}(t,\vec{y})\}=\{N_{+}(t,\vec{x}), N^{c}_{+}(t,\vec{y})\}
\end{eqnarray}
and $\gamma^{0}$-parity. An interesting aspect of the relativistic Bogoliubov transformation is that \eqref{15} implies 
\begin{eqnarray}
\{\dot{n}^{c}(t,\vec{x}), \bar{n}(t,\vec{y})\}\neq 0,
\end{eqnarray}
if one applies the Bjorken--Johnson--Low (BJL) prescription, and thus $n^{c}(x)$ is  dynamically correlated with  $\bar{n}(x)$ although $\{n^{c}(t,\vec{x}), \bar{n}(t,\vec{y})\}=0$. More interestingly, our {\it parity-doublet theorem} implies that the mass splitting of Majorana-type quasi-fermions may appear also in condensed matter physics,  if parity ($\gamma^{0}$-parity) is violated, although our analysis is strictly valid for the quadratic (mean field) approximation.   

\section{CP and related issues}  

It is well-known that Majorana neutrinos modify the CP property in electroweak interactions\cite{bilenky2}. For example, one can in principle have CP violation in the model with only two generations, although such extra CP violation is not observable in neutrino oscillation~\cite{bilenky2}.
We discuss  CP properties in the neutron oscillation using the explicit Lagrangian \eqref{3'} which is CP violating for $\alpha\neq 0$.  

To solve \eqref{3'}, we first use the Bogoliubov transformation \eqref{12}
as a change of variables, which preserves C and P. Setting $N_{+}=N$, the Lagrangian \eqref{3'}  is then written as
\begin{eqnarray}\label{16}
{\cal L}&=&(1/2)\overline{N}[i\delslash - M
-i\epsilon_{1}\sin\alpha\sin2\phi \gamma_{5}]N\nonumber\\
&+&(1/2)\overline{N^{c}}[i\delslash - M
-i\epsilon_{1}\sin\alpha\sin2\phi \gamma_{5}]N^{c}\nonumber\\
&-&(i/2)\epsilon_{1}e^{i\tilde{\alpha}}\sqrt{1-(\sin\alpha\sin2\phi)^{2}}\overline{N^{c}}N+h.c.,
\end{eqnarray}
where $\sin\tilde{\alpha}=\sin\alpha\cos2\phi/\sqrt{1-(\sin\alpha\sin2\phi)^{2}}$. After performing $N\rightarrow e^{-i\tilde{\alpha}/2}\tilde{N}$,
CP violation appears only in the parity violating mass term. A further transformation
\begin{eqnarray}\label{18}
            \frac{1}{\sqrt{2}}\left(\begin{array}{c}
            \tilde{N}(x)-i\tilde{N}^{c}(x)\\
            \tilde{N}(x)+i\tilde{N}^{c}(x)
            \end{array}\right)
            &=&\left(\begin{array}{c}
            \tilde{\varphi}_{+}(x)\\
            \tilde{\varphi}_{-}(x)
            \end{array}\right),
\end{eqnarray}
leads to
\begin{eqnarray}\label{19}
{\cal L}&=&(1/2)\overline{\tilde{\varphi}}_{+}[i\delslash - (M+\epsilon_{1}\sqrt{1-(\tilde{\epsilon}_{1}/\epsilon_{1})^{2}})-i\tilde{\epsilon}_{1}\gamma_{5}]\tilde{\varphi}_{+}\nonumber\\
&+&(1/2)\overline{\tilde{\varphi}}_{-}[i\delslash - (M-\epsilon_{1}\sqrt{1-(\tilde{\epsilon}_{1}/\epsilon_{1})^{2}})-i\tilde{\epsilon}_{1}\gamma_{5}]\tilde{\varphi}_{-},\nonumber
\end{eqnarray}
with $\tilde{\epsilon}_{1}\equiv\epsilon_{1}\sin\alpha\sin2\phi$, or after a suitable chiral transformation, one obtains a pair of Majorana fermions:
\begin{eqnarray}\label{20}
{\cal L}&=&(1/2)\bar{\varphi}_{+}(x)[i\delslash - M_{+}]\varphi_{+}(x)+(1/2)\bar{\varphi}_{-}(x)[i\delslash - M_{-}]\varphi_{-}(x),
\end{eqnarray}
where $\varphi_{\pm}=e^{i\theta_{\pm}\gamma_{5}}\tilde{\varphi}_{\pm}$ with
\begin{eqnarray}\label{phase}
(M\pm \epsilon_{1}\sqrt{1-(\tilde{\epsilon}_{1}/\epsilon_{1})^{2}})+i\tilde{\epsilon}_{1} \gamma_{5}\equiv M_{\pm}e^{2i\theta_{\pm}\gamma_{5}},
\end{eqnarray}
and
\begin{eqnarray}\label{21}
M_{\pm}=\left([M \pm \epsilon_{1}\sqrt{1-(\tilde{\epsilon}_{1}/\epsilon_{1})^{2}}]^{2}+(\tilde{\epsilon}_{1})^{2}\right)^{1/2}.
\end{eqnarray}
This mass formula covers the special cases \eqref{4} and \eqref{5} by choosing $\alpha=0$ and $\alpha=-\pi/2$, respectively. 

To analyze the possible effects of CP violation, we first  look at the parity violating mass term for the Dirac fermion in \eqref{16}, which is written
\begin{eqnarray}
{\cal L}_{mass}\simeq -(\epsilon_{1}\epsilon_{5}/M)\sin\alpha \bar{n}(x)i\gamma_{5}n(x),
\end{eqnarray}
in the leading order in $\epsilon_{1}$ and $\epsilon_{5}$. This term, which is used to evaluate the neutron electric dipole moment, should be added to the contributions from other sources such as the QCD $\theta$-vacuum.

To examine the more direct effects of CP violation in the oscillation 
amplitude we write the exact solution of the neutron field using \eqref{12}, \eqref{18} and \eqref{19},
\begin{eqnarray}\label{23}
n(x)
&=&(1/\sqrt{2})[\cos\phi e^{-i\tilde{\alpha}/2}+\gamma_{5}\sin\phi  e^{i\tilde{\alpha}/2}]e^{-i\theta_{+}\gamma_{5}}\varphi_{+}(x)\nonumber\\
&+&(1/\sqrt{2})[\cos\phi e^{-i\tilde{\alpha}/2}-\gamma_{5}\sin\phi  e^{i\tilde{\alpha}/2}]e^{-i\theta_{-}\gamma_{5}}\varphi_{-}(x),\nonumber\\
n^{c}(x)
&=&(-i/\sqrt{2})[\cos\phi e^{i\tilde{\alpha}/2}-\gamma_{5}\sin\phi  e^{-i\tilde{\alpha}/2}]e^{-i\theta_{-}\gamma_{5}}\varphi_{+}(x)\nonumber\\
&+&(i/\sqrt{2})[\cos\phi e^{i\tilde{\alpha}/2}+\gamma_{5}\sin\phi  e^{-i\tilde{\alpha}/2}]e^{-i\theta_{+}\gamma_{5}}\varphi_{-}(x).
\end{eqnarray}
We choose the P and C transformation laws of the mass eigenstates \eqref{20} consistent with the doublet representation of $\gamma^{0}$-parity as in \eqref{yy}~\cite{note}
\begin{eqnarray}\label{24}
\varphi^{p}_{\pm}(x)= \gamma^{0}\varphi_{\mp}(t,-\vec{x}), \ \ \varphi^{c}_{\pm}(x)= \pm i\varphi_{\mp}(x),
\end{eqnarray}
which induce $n\rightarrow \gamma^{0}n$ and $n^{c}\rightarrow -\gamma^{0}n^{c}$, and $n\rightarrow n^{c}$ in \eqref{23}, respectively, for the vanishing CP breaking parameter $\alpha=0$. The CP symmetry expressed by 
\begin{eqnarray}
\varphi^{cp}_{\pm}(x)=\pm i\gamma^{0}\varphi_{\pm}(t,-\vec{x})
\end{eqnarray}
does not induce the CP transformation of $n$ and $n^{c}$ for $\alpha\neq 0$, which shows the CP
breaking in \eqref{23}; to be precise, the CP transformation of $\varphi_{\pm}(x)$ in the expression of $n(x)$ does not lead to $-\gamma^{0}n^{c}(t,-\vec{x})$. We regard \eqref{23} as an analogue of the "mixing matrix" between the flavor eigenstates $(n, n^{c})$ and mass eigenstates $(\varphi_{+}, \varphi_{-})$, and we test the above CP breaking in the neutron oscillation.

One can then evaluate the oscillation probability amplitude $A=A(n(\vec{p})\rightarrow n^{c}(\vec{p});t)$ by
\begin{eqnarray}
A=\langle n^{c}_{R}(\vec{p}); 0| n_{R}(\vec{p}); t \rangle 
+\langle n^{c}_{L}(\vec{p}); 0| n_{L}(\vec{p}); t \rangle,
\end{eqnarray}
where we used the chirally projected states to take care of the $\gamma^{5}$ appearing in the above solution.
Thus,
\begin{eqnarray}
A&=&\frac{i}{2}[\cos^{2}\phi e^{-i\tilde{\alpha}}-\sin^{2}\phi  e^{i\tilde{\alpha}}]\nonumber\\
&\times&
\Big\{e^{-i\theta-iE_{+}t}\langle \varphi_{R,+}(\vec{p},0)|\varphi_{R,+}(\vec{p},0)\rangle-e^{i\theta-iE_{-}t}\langle \varphi_{R,-}(\vec{p},0)|\varphi_{R,-}(\vec{p},0)\rangle\nonumber\\
&&+e^{i\theta-iE_{+}t}\langle \varphi_{L,+}(\vec{p},0)|\varphi_{L,+}(\vec{p},0)\rangle-e^{-i\theta-iE_{-}t}\langle \varphi_{L,-}(\vec{p},0)|\varphi_{L,-}(\vec{p},0)\rangle\Big\},
\end{eqnarray}
where we defined $\theta=\theta_{+}-\theta_{-}$. We now note
\begin{eqnarray}
\langle \varphi_{R,+}(\vec{p},0)|\varphi_{R,+}(\vec{p},0)\rangle&=&\langle \varphi_{L,+}(\vec{p},0)|\varphi_{L,+}(\vec{p},0)\rangle=\frac{1}{2}\langle \varphi_{+}(\vec{p},0)|\varphi_{+}(\vec{p},0)\rangle
\end{eqnarray}
and similarly for $\varphi_{-}(\vec{p},0)$ using the $i\gamma^{0}$-parity invariance of free Majorana equations, namely, $\varphi^{p}_{R,+}(x)=i\gamma^{0}\varphi_{L,+}(t,-\vec{x})$. Using the normalization of states $\langle \varphi_{+}(\vec{p},0)|\varphi_{+}(\vec{p},0)\rangle=\langle \varphi_{-}(\vec{p},0)|\varphi_{-}(\vec{p},0)\rangle$,
we find the amplitude
\begin{eqnarray}
A
&=&[\cos^{2}\phi e^{-i\tilde{\alpha}}-\sin^{2}\phi  e^{i\tilde{\alpha}}]\cos\theta\sin\left(\frac{1}{2}\Delta E t\right)e^{-i\bar{E}t}\langle \varphi_{+}(\vec{p},0)|\varphi_{+}(\vec{p},0)\rangle,
\end{eqnarray}
with $\Delta E=E_{+}-E_{-}$ and $\bar{E}=(E_{+}+E_{-})/2$.
Adopting $|\langle \varphi_{+}(\vec{p},0)|\varphi_{+}(\vec{p},0)\rangle|^{2}=1$, one obtains the oscillation probability,
\begin{eqnarray}\label{oscillation_probab}
P(n(\vec{p})\rightarrow n^{c}(\vec{p});t)
&=&(1-\sin^{2}2\phi\cos^{2}\tilde{\alpha})\cos^{2}\theta\sin^{2}\left(\frac{1}{2}\Delta E t\right).
\end{eqnarray}
By recalling  the definitions of  $\sin\tilde{\alpha}$ in \eqref{16} and $\theta_{\pm}$ and $M_{\pm}$ in \eqref{phase},
the CP transformation, which  is equivalent to $\alpha\rightarrow-\alpha$, corresponds to 
\begin{eqnarray}
\tilde{\alpha}\rightarrow -\tilde{\alpha}, \ \ \ 
\theta=\theta_{+}-\theta_{-}\rightarrow -\theta,
\end{eqnarray}
and the above oscillation probability \eqref{oscillation_probab} and the energy difference
$\Delta E=\sqrt{\vec{p}^{2}+M^{2}_{+}}-\sqrt{\vec{p}^{2}+M^{2}_{-}}$
are all invariant. Although $\alpha$ modifies the magnitudes of $\Delta E$ (and thus the oscillation time) and probability $P$ themselves, we do not regard these modifications as a manifestation of CP violation in oscillation which is typically expressed by $P(n(\vec{p})\rightarrow n^{c}(\vec{p});t)\neq P(n^{c}(\vec{p})\rightarrow n(\vec{p});t)$. 
We observe no direct CP violation in the neutron oscillation in vacuum.    

Note that we discussed CP by looking at only the neutron sector assuming that the flavor degree ($(p,\ n)$ multiplet structure, for example) is fixed by the baryon number conserving sector of the full model, unlike the neutrino oscillation where a combination of neutrino and charged-lepton mixing matrices is analyzed.  See also~\cite{berezhiani,ft, nelson} for related analyses. 
  
The possible CPT violation in the hadron sector appears to be very small as is indicated by both experimental limit $|m_{K}-m_{\bar{K}}|<0.44\times 10^{-18}$ GeV~\cite{particledata}  and a recent model study within an extension of the Standard Model~\cite{fujikawa-tureanu}. However, we mention that the relevant mass scale of the neutron oscillation is estimated at $\epsilon_{1}\leq 6\times 10^{-24}$ eV and $n-\bar{n}$ mass splitting itself is constrained to be $\leq 10^{-15}$ eV to observe the oscillation~\cite{mohapatra2}. These values are not much different from the estimated mass difference $\sim 10^{-20}$ eV of the electron and positron induced by the possible CPT breaking that is required to explain the small ($2\sigma$) mass difference of the observed sun neutrino and reactor antineutrino as really arising from the Lorentz invariant CPT breaking~\cite{fujikawa-tureanu}. See also a recent paper~\cite{gardner}.
 
In conclusion, we have clarified the full physical contents of the model of neutron oscillation \eqref{3}. It has been shown that the basic notions such as the parity-doubling of Majorana fermions and the Bogoliubov transformation play main roles in the analysis of the model \eqref{3} which, as a relativistic analogue of BCS theory, is in turn suggestive of new possibilities in condensed matter physics. 

\subsection*{Acknowledgments}
We thank Masud Chaichian for very helpful discussions. This work is supported in part by the Vilho, Yrj\"o and Kalle V\"ais\"al\"a Foundation. The support of the Academy of Finland under the
Projects no. 136539 and 272919 is gratefully acknowledged.


\begin{thebibliography}{99}
\bibitem{note}
We assume that the free Majorana particles in Minkowski space respect the full
Lorentz symmetry, including P and T and thus C, due to CPT.
\end{thebibliography}
\end{document}